\documentclass[12pt]{iopart}

\usepackage{graphicx}
\usepackage{siunitx}

\usepackage{iopams}
\expandafter\let\csname equation*\endcsname\relax
\expandafter\let\csname endequation*\endcsname\relax
\usepackage{amsmath}
\usepackage{amsfonts}
\usepackage{amssymb}
\usepackage{bbold}
\usepackage{color}
\usepackage{bm}
\newcommand{\<}{\left\langle}
\renewcommand{\>}{\right\rangle}
\renewcommand{\(}{\left(}
\renewcommand{\)}{\right)}

\newcommand{\vect}[1]{\bm{#1}}
\newcommand{\mc}{\mathcal}
\newcommand{\ex}[1]{\mathrm{e}^{#1}}

\usepackage[breaklinks=true,colorlinks=true,linkcolor=blue,urlcolor=blue,citecolor=blue]{hyperref}

\newcommand{\rev}[1]{\textcolor{black}{#1}}

\begin{document}

\title[]{Quantifying configurational information for a stochastic particle in a flow-field}

\author{Evelyn Tang$^1$ and Ramin Golestanian$^{1,2}$}
\address{$^1$ Max Planck Institute for Dynamics and Self-Organization (MPIDS), 37077 G\"ottingen, Germany}
\address{$^2$ Rudolf Peierls Centre for Theoretical Physics, University of Oxford, Oxford OX1 3PU, United Kingdom}

\ead{ramin.golestanian@ds.mpg.de}
\vspace{10pt}
\begin{indented}
\item[]\today
\end{indented}

\begin{abstract}
Flow-fields are ubiquitous systems that are able to transport vital signalling molecules necessary for system function. While information regarding the location and transport of such particles is often crucial, it is not well-understood how to quantify the information in such stochastic systems.  Using the framework of nonequilibrium statistical physics, we develop theoretical tools to address this question. We observe that rotation in a flow-field does not explicitly appear in the generalized potential that governs the rate of system entropy production. Specifically, in the neighborhood of a flow-field, rotation contributes to the information content only in the presence of strain -- and then with a comparatively weaker contribution than strain and at higher orders in time. Indeed, strain and especially the flow divergence, contribute most strongly to transport properties such as particle residence time and the rate of information change. These results shed light on how information can be analyzed and controlled in complex artificial and living flow-based systems.
\end{abstract}

\submitto{\NJP}

\section{Introduction}

Fluid flows are present in diverse settings from microfluidic devices to solid-state and living systems, and they often transport small particles that experience stochastic dynamics and motion. Information regarding the location and transport of these particles can be vitally crucial for the function of the system. Living organisms are governed or regulated by signaling molecules in diverse situations, from the response of a slime mold to a nutrient droplet \cite{Alim5136} or in the healthy development of mammals \cite{HIROKAWA200633}. Recently, complex flow patterns have been measured in brain ventricles \cite{Faubel176}, which contain guidance molecules that are able to cue the migration and development of young neurons \cite{Sawamoto629}. Hence, it is of interest to quantify the information that is carried by a small particle in a flow-field \cite{PhysRevLett.100.178302,PhysRevE.79.040102}.

\begin{figure*}
  \centering
  \includegraphics[width=0.6\textwidth]{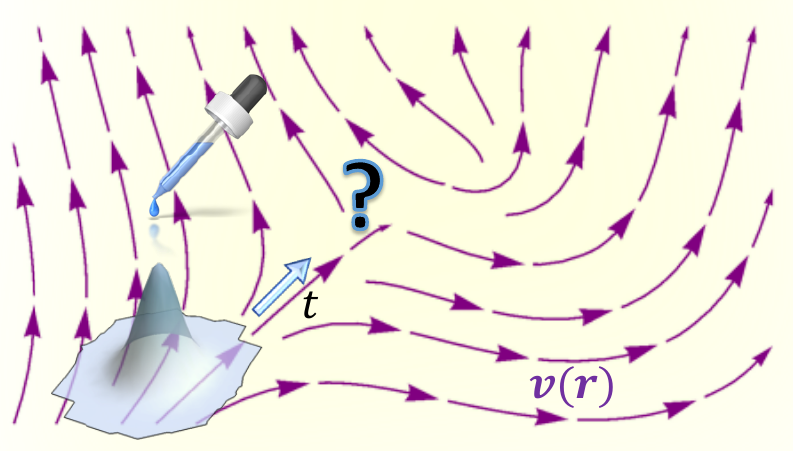}
  \caption{What is the information content after time $t$ or particle residence time, for a stochastic particle in a complex field? The answer is determined by an interplay between diffusion and advection along the flow trajectories $\vect{v}(\vect{r})$.} \label{fig:schematic}
\end{figure*}

Much progress has been made in recent years regarding how to quantify the information content in small fluctuating systems. This stems from the rigorous formulation of stochastic processes and thermodynamics of information and entropy production \cite{sekimoto,Seifert_2012,doi:10.1146/annurev-conmatphys-062910-140506,Esposito_2011,PhysRevLett.102.250602,PhysRevX.9.021009,Parrondo2015}, together with the experimental detection and manipulation of these properties in various systems including colloids \cite{Toyabe2010,C6SM00923A}, active \cite{Krishnamurthy2016,Bauerle2018} and living matter \cite{ loos2019nonreciprocal,10.1371/journal.pcbi.1003974,Gladrow2019,RibezziCrivellari2019}. These questions, however, have so far not been explored in the ubiquitous system of flow-fields in which stochastic particles are transported (see Fig. \ref{fig:schematic}).  

To answer these questions, we develop a general expression for the the rate of change of information content of a stochastic particle in a flow-field, which features contributions from the flow characteristics as well as the system fluctuations. Previous work on thermodynamics in flow-fields has focused on changes in local particle conformation \cite{PhysRevLett.100.178302,PhysRevLett.98.180603} but not on the effects of advection. The joint effects of diffusion and advection have been studied for inertial particles or regular flows such as shear or strain \cite{Taylor,RevModPhys.89.025007,RevModPhys.73.913,BENZI20092003}. Here, we develop a formalism for generic flows applicable to complex fields or realistic scenarios.

\rev{The effect of anti-symmetric or non-reciprocal terms has been a topic of great interest lately in soft and active matter systems, as they can qualitatively change the behaviour of the system in dramatic ways \cite{PhysRevLett.123.018101,loos2019nonreciprocal,Saha_2019,vitelli2020nonreciprocal,PhysRevLett.112.068301,PhysRevLett.104.178103}. Entropy production in particular has typically been identified with the part of the stochastic action which determines the weight of trajectories that is odd under time reversal \cite{Maes_odd}. In our analysis, }we find that the rotational component of a flow-field does not explicitly appear in the generalized stochastic chemical potential that governs system entropy production. 

We analyze information as the difference in entropy \cite{doi:10.1002/j.1538-7305.1948.tb01338.x,Cover:2006:EIT:1146355} between a particle in a flow-field and a particle undergoing free diffusion. In a neigborhood of a flow, the rotational component only alters the rate of information change in the presence of a strain component. Even when both strain and rotation are present, rotation provides a quantitatively weaker contribution as compared to the strain, and only at higher orders in time. Indeed, strain and especially the flow divergence, contribute most strongly to transport properties such as particle residence time and the rate of information change.

We demonstrate a wide range of possible implementations of our formalism through the study of a flow field in an arbitrary neighborhood. This allows the calculation of the change in information content and residence time scale for various geometries and flow-fields. For instance, we uncover a mechanism for retaining a particle for a longer time than diffusion would typically permit. Our results allow the quantification of information and transport properties for generic flows, which can be applied in various contexts including experimentally measured fields. 

\section{Entropy production for a particle in a flow-field}

We consider a particle with diffusion coefficient $D$ that undergoes stochastic motion in a $d$-dimensional position space under the influence of a flow-field $\vect{v}(\vect{x})$, and is characterized by the probability distribution $\mc{P}(\vect{x},t)$. To analyze the information content, we use the system (Shannon) entropy $S=-\int d^d\vect{x} \,\mc{P}\ln\mc{P}=\langle s \rangle$, where $s\equiv-\ln\mc{P}$ is the stochastic entropy of the system from a given finite trajectory \cite{PhysRevLett.95.040602}. This is an appropriate measure as it quantifies how spread out the distribution is: when the distribution is sharply peaked, the entropy is low as one can reliably locate the particle. 

To provide some intuition, we can calculate the rate of entropy production for a freely diffusive particle, which has the probability distribution
\begin{math}
\mc{P}_\textrm{D}(\vect{x},t)=\frac{1}{(4\pi Dt)^{d/2}}\exp\(-\frac{\vect{x}^2}{4Dt}\)
\end{math}. Strikingly, the rate of entropy production for this particle is simply
\begin{eqnarray}
\dot{S}_\textrm{D}(t)=\frac{d}{2t},\label{eq:sdotdiff}
\end{eqnarray} 
depending only on the system dimension $d$ and time $t$ but not the diffusion coefficient $D$. We see that the rate $\dot{S}_\textrm{D}(t)$ has a singularity at $t=0$ as the particle shifts from being strictly localized to a diffusive process. This rate then decreases with $1/t$ as the particle spreads out, to reach $0$ as the particle reaches a uniform distribution.

In the presence of a flow-field, there is an interplay between diffusion and advection. To identify the unique contributions from different components of the flow, we use the Helmholtz-Hodge decomposition to represent the vector field in terms of conservative and rotational components. This can be written as $\vect{v}(\vect{x})=-D\nabla \Phi+\vect{w}(\vect{x})$, where $\nabla \cdot \vect{w}=0$.

By invoking the continuity equation $\dot{\mc{P}}+\nabla\cdot\vect{J}=0$, where the flux is given as $\vect{J}= \vect{v}(\vect{x})\mc{P}(\vect{x},t)-D\nabla\mc{P}(\vect{x},t)$, we find a closed-form expression for the rate of entropy production as (see Appendix A for details)
\begin{eqnarray}
\dot{S}(t)&=&-D\langle\nabla^2 m\rangle,\label{eq:entprod}
\end{eqnarray}
where $m\equiv \Phi +\ln\mc{P}$ plays the role of a stochastic generalized chemical potential, which is spatially uniform in equilibrium. We thus find that system entropy production exists only when the system is manifestly out of equilibrium with a non-zero Laplacian of $m$. Remarkably, our result shows that the rotational component of the velocity, $\vect{w}$, does not explicitly appear in $m$, which governs the rate of system entropy production. 

To understand this more clearly, Eq. \eqref{eq:entprod} can be written in an alternative form as 
\begin{eqnarray}
\dot{S}(t)=\langle\nabla\cdot \vect{v}\rangle+D\langle\big(\nabla s\big)^2\rangle, \label{eq:entprod-2}
\end{eqnarray}
which highlights two things. 

First, the rotational component of the flow-field drops out of the first term, since $\vect{w}$ is divergence-free. When studying specific probability distributions for particles in a neighborhood, we will see that the rotational component only contributes to the system entropy production when strain is present, and then only in a comparatively weaker way and at higher orders in time. Intriguingly, while Landauer showed how a variable that is odd under time-reversal (current) makes additional contributions to what one expects from minimal entropy production \cite{PhysRevA.12.636,Landauer1975}, in our regime we find that a variable that is odd under time-reversal (rotation) contributes less prominently to the entropy production.
\begin{figure*}
	\begin{centering}
		\includegraphics[width=0.8\linewidth]{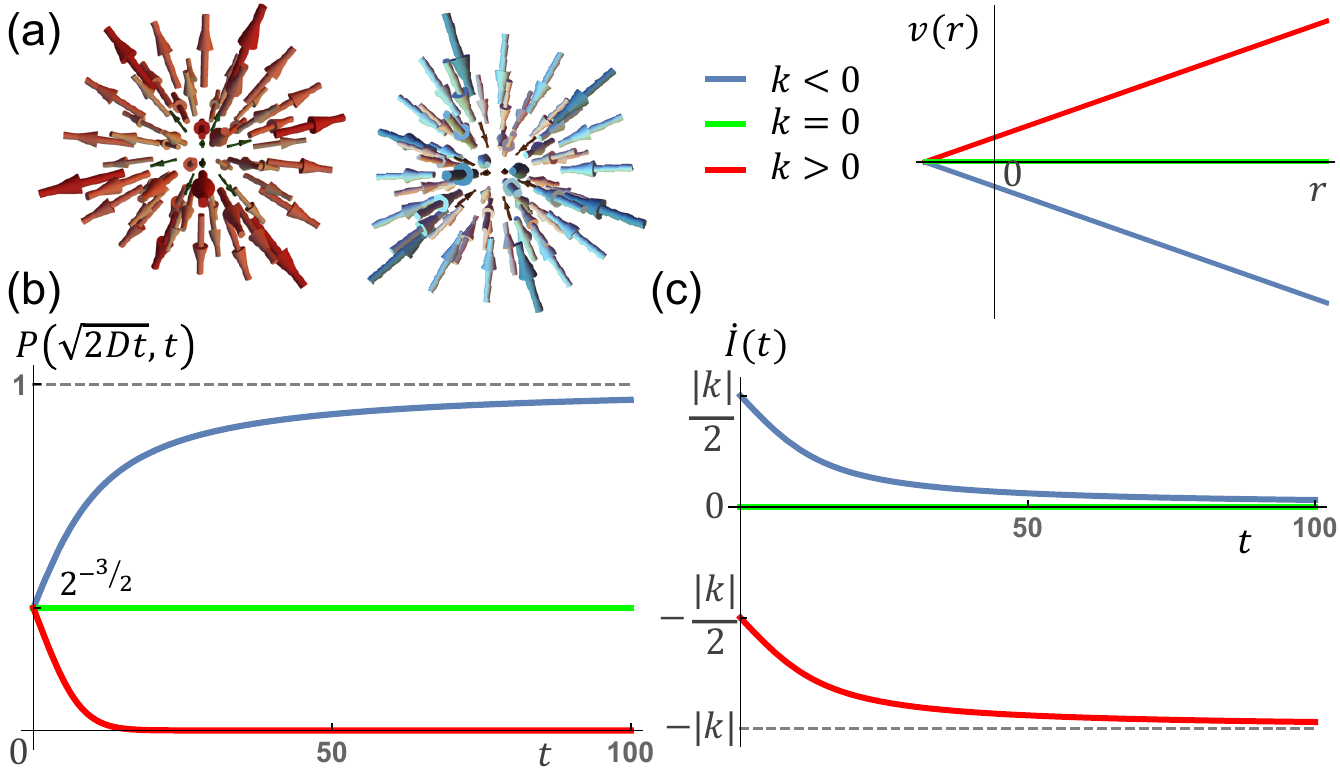}
	\caption{Transport of a particle in a divergent flow field. (a) Examples of flow fields that secrete ($k>0$, red) and absorb ($k<0$, blue) fluid (left). Flow profiles are shown; the particle starts at the origin (right). (b) The probability of observing the particle in a diffusion-limited range as a function of time. The absorbing field retains a particle for longer times (blue) compared to the the baseline set by diffusion (green), while the converse is true for the secreting field (red). (c) The rate of change of the information content as a function of time. At the beginning, the rate of change of the information content is $\dot{I}(t=0)=-\frac{1}{2}\nabla\cdot\vect{v}$ so is positive, and hence more informative for the absorbing field (blue) as compared to diffusion (green), and vice-versa for the secreting field (red). At long times, the absorbing field saturates to $\dot{I}(t)=0$, consistent with an effective steady-state distribution. Meanwhile, the secreting field saturates to $ \dot{I}(t)\to-k$, describing a probability distribution that keeps spreading out. Clearly, this is only physical within the length-scale in which the linear expansion remains valid.  All plots use $D=0.1$, $|k|=0.5$ and $\vect{r}_0=0.1\times(1,1,1)$.} \label{fig:div}
	\end{centering}
\end{figure*}

Second, the entropic contribution (second term) is positive definite. A direct corollary of this result---which Eq. (\ref{eq:entprod-2}) manifestly highlights---is that the configurational system entropy production is strictly non-negative when the flow-field is divergence-free, which will be the case if the flow corresponds to an incompressible fluid with no sources or sinks. 

\rev{\subsection{Total entropy production and entropy of the medium}}
\rev{We can further study the total entropy production $\dot{S}_{\textrm{tot}}(t)=\dot{S}(t)+{S}_{\textrm{m}}$, which has contributions from both the system entropy $\dot{S}(t)$ and entropy of the medium ${S}_{\textrm{m}}$ \cite{PhysRevLett.95.040602}. Following \cite{PhysRevLett.100.178302}, we observe that due to absence of external forces (conservative or non-conservative) the total entropy production is given by the entropic part only, namely
\begin{eqnarray}
\dot{S}_{\textrm{tot}}(t)
&=&D\langle\big(\nabla s\big)^2\rangle\geq0,
\end{eqnarray}
which guarantees that the total entropy production is positive-definite. Consequently, demanding consistency in the thermodynamic description requires the contribution to the medium entropy production to be 
\begin{equation}
\dot{S}_{\textrm{m}}(t)=-\langle\nabla\cdot \vect{v}\rangle.
\end{equation}
We observe that the change of entropy in the medium has an additional contribution when the fluid has a non-zero divergence. To understand this, we see that it is not possible to satisfy the continuity equation for an incompressible fluid (i.e. constant density) when $\nabla\cdot \vect{v} \neq 0$, unless there are sources and sinks of fluid (leading to addition or removal of fluid particles into or from the system) that are distributed across the medium in proportion to the value of $\nabla\cdot \vect{v}$. In addition to this contribution to the medium entropy production, there are other sources that contribute to heat production due to dissipation, such as internal friction that  is proportional to the viscosity and the square of the strain rate \cite{landau_lifshits_1959}. As we assume that the tracer particles do not modify the fluid flow, these contributions to medium entropy production do not change in time, and therefore, can be excluded from our thermodynamic accounting of entropy production.}

\subsection{Probability distribution for a particle in a flow-field}
To study the behavior of this expression in realistic flow-fields, we analyze the stochastic dynamics of a particle with trajectory $\vect{r}(t)$ in the presence of noise and advection due to a vector field $\vect{v}(\vect{r})$ (Fig. \ref{fig:schematic}). Specifically, in the local neighborhood of the origin, $\vect{r}=\vect{0}$, where the particle is located at $t=0$, the flow-field can be approximated using a Taylor expansion. Up to the first order in the expansion, this gives the Langevin equation
\begin{eqnarray}
\frac{d\vect{r}}{dt}=\vect{v}+\vect{K}\cdot\vect{r}+\sqrt{2D}\,\vect{\xi}
\end{eqnarray}
where $\vect{\xi}(t)$ represents a white noise (Gaussian random variable of unit strength), and we denote $v_i(\vect{0})=v_i$ and $\partial_jv_i(\vect{0})=K_{ij}$. This description will allow us to study how the 
stochastic dynamics of the particle depends on the local characteristics of the flow-field.

Using a path integral method \cite{ZinnJustin:2002ru}, we find that the probability for the particle to be found at a distance $\vect{x}$ away after time $t$ is (see Appendix B for details)
\begin{equation}
\mc{P}(\vect{x},t)=\frac{\exp\(-\frac{1}{4D}\left[\vect{x}-\vect{r}_{\rm d}(t)\right]\cdot \vect{M}^{-1}\cdot\left[\vect{x}-\vect{r}_{\rm d}(t)\right]\)}{(4\pi D)^{d/2}(\det \vect{M})^{1/2}},\label{eq:prob}
\end{equation}
where 
\begin{eqnarray}
\vect{r}_{\rm d}(t)
&=&(e^{\vect{K}t}-\vect{I})\cdot\vect{K}^{-1}\cdot\vect{v}\nonumber\\
\vect{M}(t)&=&\int_0^t d t_1 \, e^{\vect{K}(t-t_1)}\cdot e^{\vect{K}^{T}(t-t_1)}.\label{eq:rm}
\end{eqnarray}
It is useful to decompose $\vect{K}$ into its symmetric and antisymmetric components: $\vect{K}=\vect{E}-\vect{\Omega}$. Note that $\vect{E}\equiv\frac{1}{2}\left[(\nabla\vect{v})+(\nabla\vect{v})^{T}\right]$ is the strain rate tensor and $\vect{\Omega}\equiv\frac{1}{2}\left[(\nabla\vect{v})-(\nabla\vect{v})^{T}\right]$ is the vorticity tensor. They are the linearly varying components of the more general quantities, $D\nabla \Phi$ and $\vect{w}(\vect{x})$ respectively, that were defined earlier.
\begin{figure*}
\begin{centering}
	\includegraphics[width=0.7\linewidth]{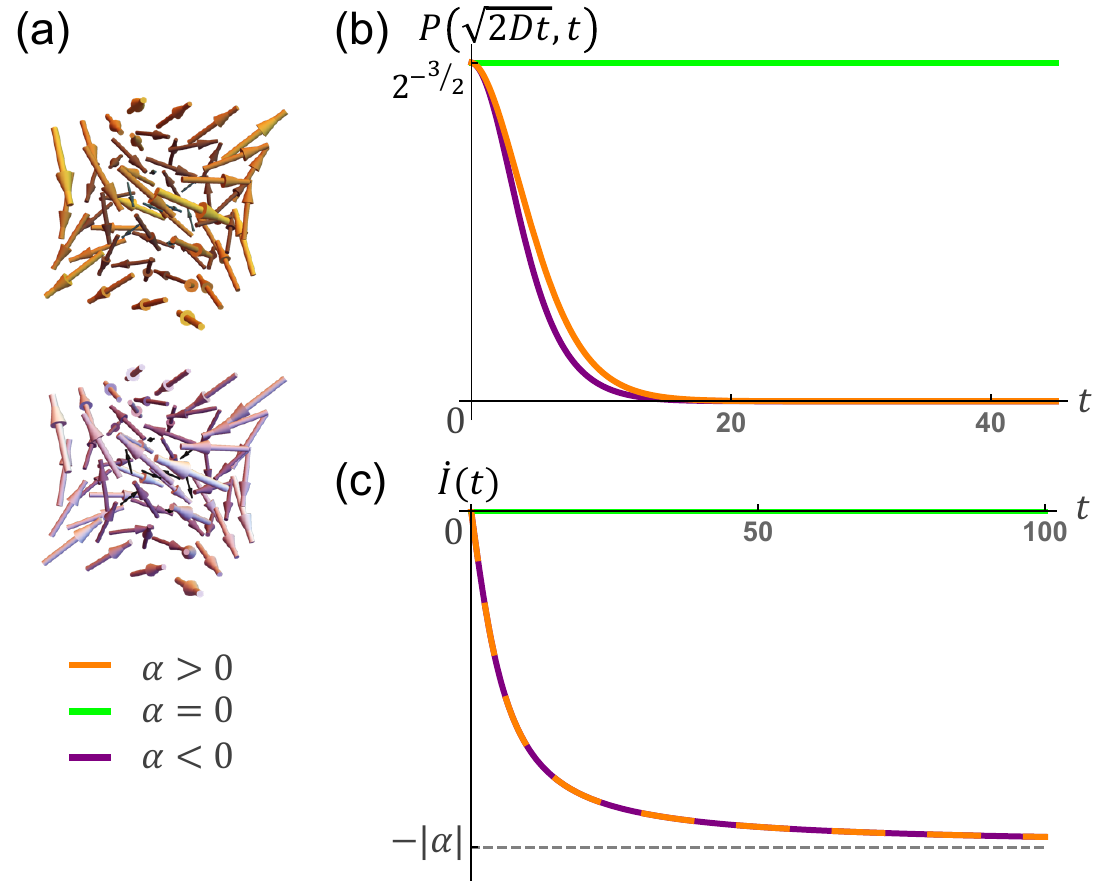}
	\caption{(a) We examine flows that have both strain and vorticity, where the transport and information properties are predicted to be dominated by the strain and not vorticity. (b)  The probability of observing the particle in a diffusion-limited range as a function of time. We observe a shorter residence time for both signs of $\alpha$ (orange and purple lines) as compared to the baseline set by diffusion (green). There is a slightly faster drop-off that depends on the direction of the strain rate $\vect{E}$ but not on the direction of vorticity $\vect{\Omega}$. 
		(c) The rate of change of the information content has no contribution at $t=0$ as there is zero divergence. The particle becomes increasingly delocalized with time as compared to diffusion, before saturating at long times to $-|\alpha|$. It is agnostic to the sign of $\alpha$ (orange and purple lines lie on top of each other). All plots again use $D=0.1$ and $\vect{r}_0=0.1\times(1,1,1)$, with $|\alpha|=0.5$.} \label{fig:vor}
		\end{centering}
\end{figure*}

\section{Rate of information change and its dependence on flow properties}

We can now calculate the average stochastic entropy for this probability distribution  
and find that Eq. \eqref{eq:entprod} gives the \rev{system} entropy production as
\begin{eqnarray}
\dot{S}(t)&=&\nabla\cdot\vect{v}+\frac{1}{2}\tr\(\vect{M}^{-1}\).\label{eq:sdot}
\end{eqnarray}
Note that this result \rev{depends only on the spatial derivative of the flow-field, and is independent of the diffusion coefficient $D$. In contrast, the entropy production of the medium can contain system-specific coefficients such as the viscous stress and fluid density \cite{landau_lifshits_1959}.}

\rev{While the expressions above are} valid for all times, we can gain intuition about the short time behavior of the result, using a series expansion
\begin{eqnarray}
\vect{M}(t)
&=&t\{\vect{I}+\vect{E}t+\frac{1}{3}\(2\vect{E}^2-[\vect{\Omega},\vect{E}]\)t^2\nonumber\\&&+\frac{1}{12}\(4\vect{E}^3+3[\vect{E}^2,\vect{\Omega}]+2[\vect{E}\vect{\Omega},\vect{\Omega}]\)t^3\nonumber\\&&+\frac{1}{60}(8\vect{E}^4+7[\vect{E}^3,\vect{\Omega}]-3[\vect{E}\vect{\Omega}\vect{E},\vect{E}]+6[\vect{E}^2\vect{\Omega},\vect{\Omega}]\nonumber\\&&+4\vect{E}\vect{\Omega}[\vect{\Omega},\vect{E}]+4[\vect{E},\vect{\Omega}^3])t^4+\mc{O}(t^5)\}\nonumber,
\end{eqnarray}where we denote the commutator $[\vect{S},\vect{T}] \equiv \vect{S}\vect{T}-\vect{T}\vect{S}$ for any two tensors $\vect{S}$ and $\vect{T}$, and $\vect{I}$ is the identity tensor. Substituting this in Eq. \eqref{eq:sdot} gives
\begin{eqnarray}
\dot{S}(t)&=&\frac{d}{2t}+\frac{1}{2}\,\nabla\cdot\vect{v}+\frac{t}{6}\tr\(\vect{E}^2\)\nonumber\\
&&-\frac{t^3}{90}\tr\(\vect{E}^4+2\vect{E}\vect{\Omega}[\vect{E},\vect{\Omega}]\)+\mc{O}(t^4),\label{eq:sdotexp}
\end{eqnarray}
where we use the fact that the trace of a commutator vanishes.

We can contrast this with our result for a freely diffusing particle in Eq. \eqref{eq:sdotdiff}. By comparing the differences in entropy production from a particle in a flow-field compared to a freely diffusing particle, we obtain the information change due to transport in the flow-field. Hence the rate of change of the information content in a flow-field is 
\begin{eqnarray}
\dot{I}(t)=-\big(\dot{S}(t)-\dot{S}_D(t)\big).\label{eq:idot}
\end{eqnarray}
Using Eqs. \eqref{eq:sdotexp} and \eqref{eq:sdotdiff}, we obtain
\begin{eqnarray}
\dot{I}(t)&=&-\frac{1}{2}\,\nabla\cdot\vect{v}-\frac{t}{6}\tr\(\vect{E}^2\)+\frac{t^3}{90}\tr\(\vect{E}^4+2\vect{E}\vect{\Omega}[\vect{E},\vect{\Omega}]\)\nonumber\\
&&+\mc{O}(t^4),\label{eq:infoflow}
\end{eqnarray}
which determines how the local properties of a flow-field---divergence, strain rate, and vorticity---affect the information content of tracer particles in any given region. 

We can see that vorticity makes a quantitatively weaker contribution compared to strain, and only at the third-order in time. In fact, examination of $\vect{M}(t)$ in  Eq. \eqref{eq:rm} reveals that for a field with pure vorticity $\vect{\Omega}\neq 0$ and no strain $\vect{E}=\vect{0}$, the exponentials in $\vect{M}(t)$ will cancel. In this case, $\vect{M}(t)=\vect{I}t$ as in free diffusion, and Eq. \eqref{eq:idot} predicts that the rate of change of information content will be 0, i.e. no different from that of a diffusive particle. 

Note that this short time expansion is consistent with our earlier expansion of the velocity field in a local neighborhood. Further, characterizing information by comparing differences in entropy across regions has been done in other contexts  \cite{Allahverdyan_2009,PhysRevX.4.031015}, including information content in gene expression levels \cite{Vergassola2007,doi:10.1146/annurev-conmatphys-031214-014803,Dubuis16301}.

\section{Particle transport and residence time scale}
\begin{figure*}
\centering
	\includegraphics[width=0.8\linewidth]{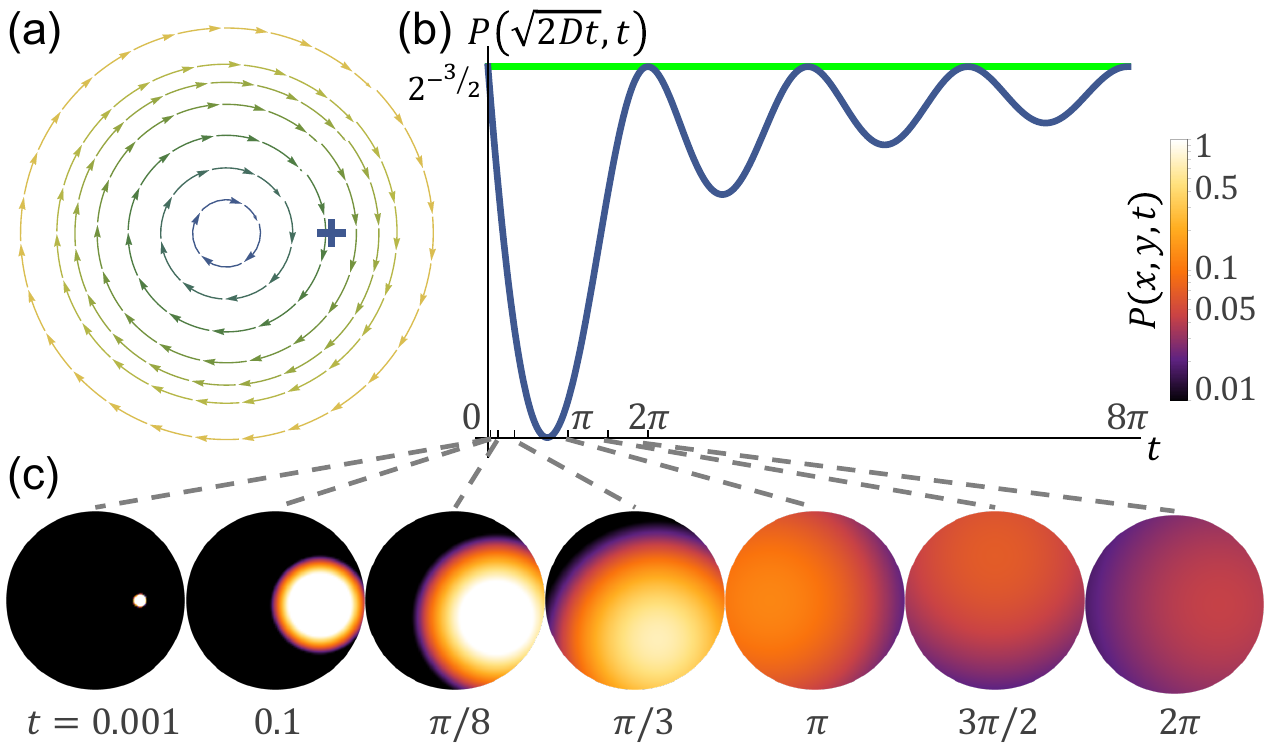}
	\caption{(a) A field  that has only vorticity $\vect{\Omega}\neq 0$ and no strain $\vect{E}=\vect{0}$. The particle begins at the blue cross to the right of the vortex center. (b) The probability $P(\sqrt{2Dt},t)$ is plotted for that point (blue), where we see oscillations with a period of $2\pi/c$ from the baseline value of diffusion (green). (c) The probability distribution $ \mc{P}(x,y,t)$ is plotted at various time points within a period, showing oscillatory motion around the vortex center that decays due to diffusion. Fields with pure vorticity such as this one have no change in their information content as compared with free diffusion. These plots again use $D=0.1$, with $c=1$ and $\vect{r}_0=(0.5,0,0)$.} \label{fig:2dvortex}
\end{figure*}

Before illustrating these results with specific cases, we analyze the transport observables such as residence time of a particle before it is washed away by the flow-field. This is given by the probability to observe the tracer particle in a region of $\sigma^d$ around the origin after time $t$, defined as $P(\sigma,t)=\int d^d\vect{x}\,\ex{-\frac{\vect{x}^2}{2\sigma^2}}\mc{P}(\vect{x},t)$. Using our solution in Eq. \eqref{eq:prob}, we find
\begin{eqnarray}
P(\sigma,t)&=&\(\frac{\sigma^2}{2D}\)^{d/2}\frac{\exp\(-\frac{1}{4D}\vect{r}_d(t)\cdot[\vect{M}+\frac{\sigma^2}{2D}\vect{I}]^{-1}\cdot\vect{r}_d(t)\)}{\det [\vect{M}+\frac{\sigma^2}{2D}\vect{I}]^{1/2}}.\nonumber\\
\label{eq:aftert}
\end{eqnarray} 
We choose $\sigma^2=2Dt$, the length scale set by diffusion, in order to continue comparing the behavior of our particle in a flow to that of free diffusion. Using a short time expansion, we find
\begin{eqnarray}
P(\sqrt{2Dt},t)&\approx&\frac{1}{2^{d/2}}\,\exp[-\frac{1}{4}\(\nabla\cdot\vect{v}+\frac{\vect{v}^2}{2D}\)t\nonumber\\&&-\(\frac{\vect{v}\cdot\vect{E}\cdot\vect{v}}{16 D}+\frac{5}{48}\tr(\vect{E}^2)\)t^2 + \mc{O}(t^3)].\nonumber\\\label{eq:restimediff}
\end{eqnarray}
Keeping only the lowest order term, we have
\begin{equation}
P(\sqrt{2Dt},t)\approx \frac{1}{2^{d/2}}\exp\(-\frac{t}{2\mc{T}_r}\), \label{p-sigma-t}
\end{equation}
where 
\begin{equation}
\mc{T}_r^{-1}=\frac{1}{2}\(\nabla\cdot\vect{v}+\frac{\vect{v}^2}{2D}\), \label{Tr-def}
\end{equation}
gives the time scale for a particle to remain in a region set by diffusion. 

If instead we choose the region of interest to be in the close vicinity of the origin, i.e. $\sigma^2\ll2Dt$, then we obtain
\begin{eqnarray}
P(\sigma\ll\sqrt{2Dt},t)
&\approx&\(\frac{\sigma^2}{2Dt}\)^{d/2}\exp\(-\frac{t}{\mc{T}_r}\),\label{p-sigma-lt-t}
\end{eqnarray}
which is controlled by the same residence time scale.

\section{Illustration through specific cases}
\subsection{Fields with sources or sinks}
These results predict that a divergent field will modify the transport behavior strongly from what is expected in the case of diffusion. In particular, Eq. \eqref{Tr-def} shows that a field with negative divergence, e.g. in an absorbing tissue, can retain a particle for much longer times. We verify this in an example where fluid is secreted or absorbed at rate $k$ as described by the flow field 
	$\vect{v}(\vect{r})=\frac{k}{3} (\vect{r}-\vect{r}_0)$ which can be seen in Fig. \ref{fig:div}a. Figure \ref{fig:div}b demonstrates that an absorbing field retains a particle for longer times as compared to diffusion, while the opposite is observed for a secreting field. 

Furthermore, we can use Eq. \eqref{eq:rm} to calculate the Gaussian covariance as $\vect{M}(t)=\frac{3}{2k}[\exp\left(\frac{2kt}{3}\right)-1]\vect{I}$.  As $t\to\infty$, this covariance has different asymptotic behaviors depending on the sign of $k$. In a secreting field with $k>0$, $\vect{M}(t)$ diverges exponentially at long times, which means that the particle becomes dispersed into the fluid and $P(\sqrt{2Dt},t)\to0$ (red line in Fig. \ref{fig:div}b). Instead, for an absorbing field with $k<0$, it saturates to a constant at long times as $\vect{M}(t)\to\frac{3}{2|k|}\vect{I}$. Hence, the probability for this case approaches unity ($P(\sqrt{2Dt},t)\to1$) (blue line in Fig. \ref{fig:div}b), a manifestation of the localization of the particle.

Using Eq. \eqref{eq:infoflow}, we note that the change of information content has an initial value of $\dot{I}(0)=-\frac{1}{2}\nabla\cdot\vect{v}$. This is positive for the absorbing field $k<0$, which is more localized as compared to diffusion and hence more informative. The converse is true for the secreting field (see Fig. \ref{fig:div}c). At long times, $\frac{1}{2}\tr\(\vect{M}^{-1}\)$ goes to $0$ when $k<0$ and to $-|k|$ when $k>0$. Hence, Eqs. \eqref{eq:sdot} and \eqref{eq:idot} show that the absorbing field saturates to $\dot{I}(t)=0$, consistent with an effective steady-state distribution. However, the secreting field saturates to $ \dot{I}(t)\to-k$, describing a probability distribution that spreads out at a faster rate than diffusion. Clearly, this is only physical within the length-scale in which the linear expansion remains valid, $l \sim\frac{|K_{jk}|}{|\partial_iK_{jk}|}$, and within $t\sim \frac{1}{k}\ln\frac{l}{r_0}$ before boundary effects come into play. 

\subsection{Fields with both strain and vorticity}
In fields that have both strain and vorticity components, Eqs. \eqref{eq:infoflow} and \eqref{eq:restimediff} predict that both the change in information content and particle residence time will be dominated by strain and not vorticity. To see this, we examine flow-fields with both these components in 3d, while retaining linear flow profiles and parameters similar to the previous examples (see Fig. \ref{fig:vor}a): 
\begin{eqnarray}
\vect{v}(\vect{r})&=&\alpha\left[(z-z_0)\hat{\vect{e}}_{x}+(x-x_0)\hat{\vect{e}}_{y}+(y-y_0)\hat{\vect{e}}_{z}\right].\label{eq:vor3d}
\end{eqnarray}

As expected from Eq. \eqref{eq:restimediff}, the residence time in these fields is shorter than in free diffusion (see Fig. \ref{fig:vor}b). The cases with positive and negative values of $\alpha$ have very similar features, although there is a slightly faster drop-off that depends on the direction of the strain rate $\vect{E}$ but not on the direction of vorticity $\vect{\Omega}$. 

Eq. \eqref{eq:infoflow} predicts no initial contribution at $t=0$ in the change of information content since these fields are divergence-free. Instead, the leading term in time is linear and negative, $\dot{I}(t)\sim -\frac{\alpha^2}{4}t$, hence these fields become increasingly delocalized with time as compared to diffusion (see Fig. \ref{fig:vor}c). At long times, this expression saturates in the same way to $-|\alpha|$ for both signs of $\alpha$. This is because $\frac{1}{2}\tr\(\vect{M}^{-1}\)=\frac{\alpha[2+\cosh(\alpha t)]}{\sinh(\alpha t)}$, which is even with respect to $\alpha$ (the orange and purple plots in Fig. \ref{fig:vor}c lie on top of each other). Similar to the previous case of a secreting field, $\dot{I}(t)<0$ only makes sense within the time- and length-scale of validity for the linear expansion.

\subsection{Fields with only vorticity}
As discussed in Eqs. \eqref{eq:idot}-\eqref{eq:infoflow}, a field with pure vorticity $\vect{\Omega}\neq 0,\vect{E}=\vect{0}$ has $\vect{M}(t)=\vect{I}t$ as in free diffusion, independent of $\vect{\Omega}$. In this case, the rate of change of information content is 0, i.e. no different from that of a freely diffusing particle. While $\vect{M}(t)$ no longer depends on $\vect{\Omega}$, we note that $\vect{r}_{\rm d}$ and hence the exponential term in the probability expression still depends on $\vect{\Omega}$, as can be seen from Eq. \eqref{eq:rm}. The vorticity $\vect{\Omega}\neq0$ will result in $e^{\vect{K}t}$ having complex roots and hence oscillations. 

To demonstrate this, we study a flow-field with pure vorticity (Fig. \ref{fig:2dvortex}a), which has $\vect{\Omega}\neq 0,\vect{E}=\vect{0}$:
\begin{eqnarray}
\vect{v}(\vect{r})&=&-c \, \hat{\vect{e}}_{z} \times (\vect{r}-\vect{r}_0).
\end{eqnarray}
At the origin $(x,y)=(0.5,0)$ (the blue cross in Fig. \ref{fig:2dvortex}a), the particle rotates around the vortex center and hence the probability in that region decreases before returning to the baseline set by diffusion after a full period $2\pi/c$ (the blue plot in Fig. \ref{fig:2dvortex}b). The probability density around the vortex is plotted at various times within a period to show the oscillatory motion, which simultaneously decays due to diffusion (Fig. \ref{fig:2dvortex}c). 

\begin{figure*}[t]
\centering
	\includegraphics[width=0.7\linewidth]{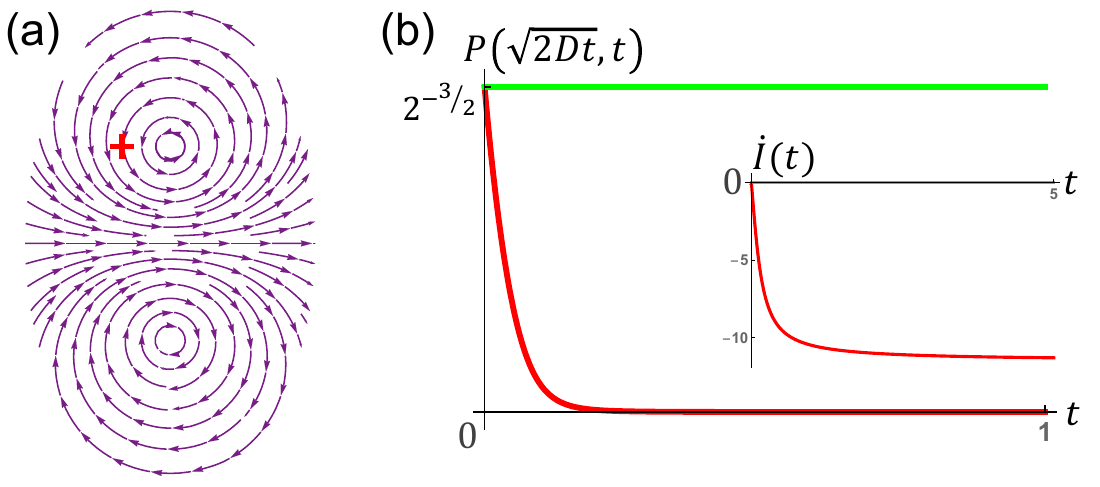}
	\caption{(a) We examine a complex field: in which two vortices with opposite vorticity overlap. Their centers are $\vect{r}_1=(0.5,0,0)$ and $\vect{r}_2=(0.5,-2,0)$ respectively from the origin (red cross). (b) This vorticity causes the probability density at the origin (red) to vanish more quickly compared to pure diffusion (green). (c) The change in information content begins from 0 and decreases with time to saturate, as we expect from flows with both vorticity and strain contributions. The plot region is an ellipse with 3 units wide and 5 units high, with similar parameters as in the previous figures.} \label{fig:twovortices}
\end{figure*}

\subsection{More complex fields}
As our formalism is applicable in the local neighborhood of a generic flow, it can be applied to various scenarios including more complex flow patterns. To illustrate this, we examine a complex field: in which two vortices with opposite vorticity $\vect{v}(\vect{r})=\pm\frac{\hat{\vect{e}}_{z} \times (\vect{r}-\vect{r}_{1,2})}{|\vect{r}-\vect{r}_{1,2}|^3}$ overlap (see Fig. \ref{fig:twovortices}). Their centers are $\vect{r}_1=(0.5,0,0)$ and $\vect{r}_2=(0.5,-2,0)$ respectively from the origin (red cross). 

In this case, we find that the change in information content begins from 0 and decreases with time to saturate, i.e. the probability distribution spreads faster than diffusion. This particle delocalization is what we would expect from flows with both anti-symmetric and symmetric contributions, similar to what we saw in Fig. \ref{fig:vor}.

\section{Discussion}
We derive analytical expressions for the rate of change of information content that explicitly illustrate its out-of-equilibrium character, through the introduction of a stochastic generalized chemical potential. We find that the leading contribution in time stems from the field divergence, and that the rotational component only makes a subleading contribution. We use this to study the information content and particle transport for a stochastic particle in a flow-field. 

In a neighborhood of a flow, vorticity only contributes to the change of information content when there is an additional strain field component, but produces oscillations in the probability density. Even when both strain and rotation are present, rotation provides a quantitatively weaker contribution as compared to the strain, and only at higher orders in time. Similarly, strain and especially the flow divergence, are found to contribute much more strongly than vorticity to transport properties such as the particle residence time.

This formalism is applicable in the local neighborhood of a generic flow and can be applied to more complex flow patterns. For instance, it can be used for experimentally-measured flow-fields in each local neighborhood, and hence within complex geometries such as those in biological tissues \cite{Faubel176}. The rich transport behavior is further characterized through an expression for the particle residence time, through which we identify a mechanism to retain a particle for longer times compared to diffusion. Our identification of the way in which local flow properties determine the information content and transport properties, can be used to design desired changes in information transmission.

\rev{Our work builds upon an earlier paper by Speck and Seifert that examine entropy production in a flow field \cite{PhysRevLett.100.178302}, which focuses exclusively on the case where $\nabla\cdot \vect{v}=0$. While our results are in agreement with Ref. \cite{PhysRevLett.100.178302} for $\nabla\cdot \vect{v}=0$, when $\nabla\cdot \vect{v}\neq0$, and therefore there are sources and sinks of fluid, we find new contributions to entropy production and qualitatively different results. Further, as we directly calculate probability distributions, we can explicitly evaluate the ensemble averages to identify dominant terms and the system behaviour, for both the system entropy and the rate of change of information. In addition, our study of the residence time in various flow fields provides another useful metric for the study of particle transport.}

\rev{Our methods employ an initial condition that is sharply localized, in order to analyze the transient behaviour of the system within the region of expansion. This approach affords us with analytical tractability, and develops predictive tools that decipher the corresponding roles of the different characteristic properties of the flow field. Our study can be complemented with computational approaches that relax some of these approximations and focus on more realistic scenarios, e.g. when the particle moves far from the initial point.}

This work opens many new directions as our analysis can be extended to include the presence of different chemical species and gradients \cite{PhysRevFluids.1.034001}, or non-conserved particle densities. 
It would also be of great interest to probe how information can create feedback loops or time-dependent control of the flow field, as well as the possibility of learning from the available information \cite{Barato_2014,PhysRevLett.120.020601}. Overall, such work will inform the transmission of information in diverse scenarios, that are relevant for a range of vital chemical and mechanical processes.

\section*{Acknowledgments}
We thank L\'{e}o McCormack, Babak Nasouri and Andrej Vilfan for helpful discussions.

\begin{appendix}

\section{Entropy production in a flow-field}
We calculate the system (Shannon) entropy for the distribution of a particle in a flow-field $\vect{v}(\vect{x})$
\begin{eqnarray}
S(t)\equiv-\int d^d\vect{x}\,\mc{P}(\vect{x},t)\ln\mc{P}(\vect{x},t),
\end{eqnarray}
which quantifies how spread out the distribution is. We use the continuity equation $\dot{\mc{P}}+\nabla\cdot\vect{J}=0$, with the flux is given by
\begin{equation}
\vect{J}= \vect{v}(\vect{x})\mc{P}(\vect{x},t)-D\nabla\mc{P}(\vect{x},t)=\mc{P} \left[\vect{v}+D \nabla s\right], \label{eq:flux}
\end{equation}
where we have used the definition $s\equiv-\ln\mc{P}$ for the stochastic entropy of the system. The rate of entropy production can now be calculated as
\begin{eqnarray}
\dot{S}(t)&=&\int d^d\vect{x}\,\nabla\cdot\vect{J}\ln\mc{P}+\int d^d\vect{x}\nabla\cdot\vect{J},\nonumber\\
&=&-\int d^d\vect{x}\,\vect{J}\cdot\nabla\ln\mc{P},\label{eq:J-grad}\\
&=&-\int d^d\vect{x}\left[\vect{v}\cdot\nabla\mc{P}+D\mc{P} \big(\nabla s\big)^2\right],\nonumber\\
&=&\langle\nabla\cdot \vect{v}\rangle+D\langle\big(\nabla s\big)^2\rangle
\end{eqnarray}
plus boundary terms, which we assume to be negligible. 

Further, we use the Helmholtz-Hodge decomposition to decompose the vector field $\vect{v}(\vect{x})=-D\nabla \Phi+\vect{w}(\vect{x})$, into its rotational and conservative components, respectively. $\nabla \cdot \vect{w}=0$. Then 
\begin{eqnarray}
\vect{J}&=&\mc{P}[\vect{w}(\vect{x})-D\nabla m],\label{eq:w-J}
\end{eqnarray}
where $m\equiv \Phi +\ln \mc{P}=\Phi -s$. Equations \eqref{eq:J-grad} and \eqref{eq:w-J} give
\begin{eqnarray}
\dot{S}(t)&=&-\int d^d\vect{x} \,[\vect{w}(\vect{x})-D\nabla m]\cdot\nabla\mc{P},\nonumber\\
&=&-D\langle\nabla^2 m\rangle = -D\langle\nabla^2 \Phi\rangle+D\langle\nabla^2 s\rangle.\nonumber
\end{eqnarray}
where we used the fact that $\nabla\cdot \vect{w}=0$.

\section{Probability distribution of a particle in an arbitrary linear flow-field}

We start with the Langevin equation
\begin{eqnarray}
\frac{d\vect{r}}{dt}=\vect{v}+\vect{K}\cdot\vect{r}+\sqrt{2D}\,\vect{\xi},
\end{eqnarray}
where we denote $v_i(\vect{0})=v_i$ and $\partial_jv_i(\vect{0})=K_{ij}$.

We can use a path-integral method to obtain the solution. First, the linear equation allows the solution to be broken into the {\it deterministic} and {\it fluctuating} contributions:
\begin{eqnarray}
\vect{r}(t)=\vect{r}_d(t)+\vect{r}_f(t),
\end{eqnarray}
where 
\begin{eqnarray}
\frac{d\vect{r}_d}{dt}=\vect{v}+\vect{K}\cdot\vect{r}_d.
\end{eqnarray}
This has the solution
\begin{eqnarray}
\vect{r}_d(t)&= &e^{\vect{K}t}\cdot\vect{r}_d(0)+\int_0^tdt_1e^{\vect{K}(t-t_1)}\cdot\vect{v},\nonumber\\
&=&e^{\vect{K}t}\cdot\vect{r}_d(0)+(e^{\vect{K}t}-\vect{I})\vect{K}^{-1}\cdot\vect{v}.
\end{eqnarray}
In the main text we have set $\vect{r}_d(0)=\vect{0}$.

Besides this, 
\begin{eqnarray}
\frac{d\vect{r}_f}{dt}=\vect{K}\cdot\vect{r}_f+\sqrt{2D}\,\vect{\xi},
\end{eqnarray}
where we set $\vect{r}_f(0)=\vect{0}$. Then

\begin{eqnarray}
&&e^{\vect{K}t}\cdot\frac{d}{dt}\left[e^{-\vect{K}t}\cdot\vect{r}_f(t)\right]=\sqrt{2D}\,\vect{\xi},\nonumber\\
&&\vect{r}_f(t)=\sqrt{2D}\int_0^tdt_1\,e^{\vect{K}(t-t_1)}\cdot\vect{\xi}(t_1),
\end{eqnarray}
so 
\begin{eqnarray}
\vect{r}(t)=\vect{r}_d(t)+\sqrt{2D}\int_0^tdt_1\,e^{\vect{K}(t-t_1)}\cdot\vect{\xi}(t_1).
\end{eqnarray}

We are interested in the probability of the particle to be found at a distance $\vect{x}$ away after time $t$, i.e.
\begin{eqnarray*}
\fl\mc{P}(\vect{x},t)=\<\delta^d\(\vect{x}-\vect{r}(t)\)\>,\nonumber\\
\fl=\int\mc{D}\vect{\xi}\,\mc{P}[\vect{\xi}]\,\delta^d\(\vect{x}-\vect{r}_d(t)-\sqrt{2D}\int_0^t dt_1\,e^{\vect{K}(t-t_1)}\cdot\vect{\xi}(t_1)\),\nonumber\\
\fl=\frac{1}{\mc{Z}}\int\mc{D}\vect{\xi} \int_{-\infty}^\infty\frac{d^d\vect{\lambda}}{(2\pi)^d}\,\exp\left[-\frac{1}{2}\int dt_1\,\vect{\xi}(t_1)^2+i\vect{\lambda}\cdot\(\vect{x}-\vect{r}_d(t)-\sqrt{2D}\int_0^tdt_1\,e^{\vect{K}(t-t_1)}\cdot\vect{\xi}(t_1)\)\right],\nonumber\\
\fl=\int_{-\infty}^\infty\frac{d^d\vect{\lambda}}{(2\pi)^d}\,e^{i\lambda\cdot[\vect{x}-\vect{r}_d(t)]}\exp\left[-D\lambda^T\cdot\vect{M}\cdot\lambda\right],\label{eq:prob2}\\
\fl=\frac{1}{(4\pi D)^{d/2}(\det \vect{M})^{1/2}}\exp\(-\frac{1}{4D}\left[\vect{x}-\vect{r}_d(t)\right]\cdot \vect{M}^{-1}\cdot\left[\vect{x}-\vect{r}_d(t)\right]\),\nonumber
\end{eqnarray*}
where $\mc{Z}=\int\mc{D}\vect{\xi} \,\exp\left[-\frac{1}{2}\int dt_1\vect{\xi}(t_1)^2\right]$ and 
\begin{eqnarray}
\vect{M}&=&\int_0^tdt_1\,e^{\vect{K}(t-t_1)}e^{\vect{K}^T(t-t_1)}.\nonumber
\end{eqnarray}

\section{Particle residence time}

We would like to analyze the residence time of a tracer particle in a particular location before it is washed away by the flow-field. This can be characterized by the probability to observe the tracer particle in a region of size $\sim\sigma^d$ after time $t$, i.e.
\begin{eqnarray}
P(\sigma,t)&=&\int d^d\vect{x}\,\ex{-\frac{\vect{x}^2}{2\sigma^2}}\mc{P}(\vect{x},t),\nonumber\\
&=&\int d^d\vect{x}\,\ex{-\frac{\vect{x}^2}{2\sigma^2}}\int_{-\infty}^\infty\frac{d^d\vect{\lambda}}{(2\pi)^d}\ex{i\vect{\lambda}\cdot[\vect{x}-\vect{r}_d(t)]}\ex{-D\vect{\lambda}^T\cdot\vect{M}\cdot\vect{\lambda}},\nonumber\\
&=&\(\frac{\sigma^2}{2D}\)^{d/2}\frac{\exp\(-\frac{1}{4D}\vect{r}_d(t)\cdot [\vect{M}+\frac{\sigma^2}{2D}\vect{I}]^{-1}\cdot\vect{r}_d(t)\)}{\det [\vect{M}+\frac{\sigma^2}{2D}\vect{I}]^{1/2}},\label{eq:aftert}
\end{eqnarray} 
where we used the expression for the probability distribution in Eq. \eqref{eq:rm}.  

Analyzing the small-time behavior of this expression too, we have
\begin{eqnarray}
\fl\vect{M}+\frac{\sigma^2}{2D}\vect{I}&=\frac{\sigma^2}{2D}\vect{I}+t\(\vect{I}+\vect{E}t+\frac{1}{3}\(2\vect{E}^2-[\vect{\Omega},\vect{E}]\)t^2+\mc{O}(t^3)\),\nonumber\\
\fl&=\(\frac{\sigma^2}{2D}+t\)\(\vect{I}+\frac{1}{\(\frac{\sigma^2}{2Dt}+1\)}\vect{E}t+\frac{1}{3}\frac{1}{\(\frac{\sigma^2}{2Dt}+1\)}\(2\vect{E}^2-[\vect{\Omega},\vect{E}]\)t^2+\frac{1}{\(\frac{\sigma^2}{2Dt}+1\)}\mc{O}(t^3)\),\nonumber\end{eqnarray}
so
\begin{eqnarray}
\fl\left[\vect{M}+\frac{\sigma^2}{2D}\vect{I}\right]^{-1}=\nonumber\\
\fl\frac{2D}{\sigma^2+2Dt}\(\vect{I}-\frac{1}{\(\frac{\sigma^2}{2Dt}+1\)}\vect{E}t-\frac{1}{3}\frac{1}{\(\frac{\sigma^2}{2Dt}+1\)}\(2\vect{E}^2-[\vect{\Omega},\vect{E}]\)t^2+\frac{1}{\(\frac{\sigma^2}{2Dt}+1\)^2}\vect{E}^2t^2+\frac{1}{\(\frac{\sigma^2}{2Dt}+1\)}\mc{O}(t^3)\).\nonumber
\end{eqnarray}

Setting $\vect{r}_d(0)=\vect{0}$, i.e. the particle begins at the origin at $t=0$,
\begin{eqnarray}
\vect{r}_d(t)&=&\((\vect{I}+\vect{K}t+\frac{1}{2}\vect{K}^2t^2)-\vect{I}\)\vect{K}^{-1}\cdot\vect{v}+\mc{O}(t^3),\nonumber\\
&=&\vect{v}t+\frac{1}{2}\vect{K}\cdot\vect{v}t^2+\mc{O}(t^3),\nonumber\\
&=&\vect{v}t+\frac{1}{2}\vect{E}\cdot\vect{v}t^2+\mc{O}(t^3),
\end{eqnarray}
where we use $\vect{v}\cdot\vect{\Omega}\cdot\vect{v}=0$.

Then the argument of the exponential term in $P(\sigma,t)$ from Eq. \eqref{eq:aftert} is
\begin{eqnarray}
\fl-\frac{1}{4D}\vect{r}_d(t)^T \cdot[\vect{M}+\frac{\sigma^2}{2D}\vect{I}]^{-1}\cdot\vect{r}_d(t)
&\approx &-\frac{1}{2(\sigma^2+2Dt)}\(\vect{v}^2t^2+\frac{\sigma^2}{\sigma^2+2Dt}\vect{v}^T\cdot\vect{E}\cdot\vect{v}t^3\)+ \mc{O}(t^3).
\nonumber\\
\end{eqnarray}

The full expression for probability (Eq. \ref{eq:aftert}) contains $\det [\vect{M}+\frac{\sigma^2}{2D}\vect{I}]$, which can be expanded as 
\begin{eqnarray}
\fl\det [\vect{M}+\frac{\sigma^2}{2D}\vect{I}]\nonumber\\
\fl=\(\frac{\sigma^2}{2D}+t\)^d\exp\left[\tr \ln\(\vect{I}+\frac{1}{\(\frac{\sigma^2}{2Dt}+1\)}\vect{E}t+\frac{1}{3}\frac{1}{\(\frac{\sigma^2}{2Dt}+1\)}\(2\vect{E}^2-[\vect{\Omega},\vect{E}]\)t^2+\frac{1}{\(\frac{\sigma^2}{2Dt}+1\)}\mc{O}(t^3)\)\right],\nonumber\\
\fl=\(\frac{\sigma^2}{2D}+t\)^d\exp\left[\frac{\tr(\vect{E})t}{\(\frac{\sigma^2}{2Dt}+1\)}+\frac{\tr(\vect{E}^2)t^2}{\(\frac{\sigma^2}{2Dt}+1\)}\(\frac{2}{3}-\frac{1}{2\(\frac{\sigma^2}{2Dt}+1\)}\)+\frac{1}{\(\frac{\sigma^2}{2Dt}+1\)}\mc{O}(t^3)\right].
\end{eqnarray}
Here we have used the fact that the inner product of a symmetric and antisymmetric matrix vanishes, i.e. $\tr(\vect{\Omega}\vect{E})=0$.

Then, it follows that
\begin{eqnarray}
\fl P(\sigma,t)\approx\(1+\frac{2Dt}{\sigma^2}\)^{-d/2}\times\nonumber\\
\fl\exp\(-\frac{1}{2\(\frac{\sigma^2}{2Dt}+1\)}\left[\(\nabla\cdot\vect{v}+\frac{1}{2D}\vect{v}^2\)t+\(\frac{\sigma^2\vect{v}^T\cdot\vect{E}\cdot\vect{v}}{2D(\sigma^2+2Dt)}+\tr(\vect{E}^2)\(\frac{2}{3}-\frac{1}{2\(\frac{\sigma^2}{2Dt}+1\)}\)\)t^2\right]\),\nonumber\\
\end{eqnarray} 
which can be used to calculate the asymptotic forms reported in the main text.

\end{appendix}

\section*{References}
\bibliography{stochasticflow}
\bibliographystyle{iopart-num}

\end{document}